\documentclass[usenatbib, letters]{mnras}
\usepackage{amssymb,amsmath}
\usepackage{graphicx}
\usepackage{xcolor}
\usepackage{multirow}
\usepackage[T1]{fontenc}
\usepackage{ae,aecompl}
\usepackage{newtxtext,newtxmath}

\def\msig{$M_{\rm BH}- \sigma$\ }
\def\nii{[N~{\sc ii}]$\lambda6583$\AA}
\def\si{[S~{\sc ii}]$\lambda6716$\AA}
\def\sii{[S~{\sc ii}]$\lambda6732$\AA}

\title[Broad H$\alpha$ variability in SDSS J0159]
{Further evidence to support a tidal disruption event in the changing-look AGN SDSS J0159}
\author[Zhang]
{Xue-Guang Zhang\thanks{Corresponding author
    Email: \href{mailto:xgzhang@njnu.edu.cn}{xgzhang@njnu.edu.cn}}\\
    School of physics and technology, Nanjing Normal University, No. 1, Wenyuan Road, 210046, P. R. China}
\date{}

\begin{document}

\pagerange{\pageref{firstpage}--\pageref{lastpage}} \pubyear{2020}

\maketitle

\label{firstpage}

\begin{abstract} 
   In this Letter, we report further evidence to support a tidal disruption event (TDE) in the known 
changing-look AGN SDSS J0159, through the unique variabilities of asymmetric broad H$\alpha$: broader but 
redder in 2010 than in 2000. Accepted the broad H$\alpha$ emission regions tightly related to accreting 
debris in a central TDE, the well-known relativistic accretion disk origination can be applied to well 
explain the asymmetric broad H$\alpha$ variabilities in SDSS J0159. Moreover, the model determined broad 
H$\alpha$ emission regions have the sizes not follow the empirical R-L relation valid in normal broad line 
AGN, but have locations basically similar to the regions of accreting debris from a central TDE in SDSS 
J0159, indicating tight connections between the broad H$\alpha$ emission materials and central TDE debris. 
Therefore, explanations of the asymmetric broad H$\alpha$ variabilities through the relativistic accretion 
disk origination provide further clues to support a central TDE in SDSS J0159.
\end{abstract}


\section{Introduction}

    TDEs (Tidal Disruption Events) have been studied in detail for more than four decades \citep{re88, 
lu97, gs12, gr13, gm14, ht16, wy18, tc19}, with accreting fallback debris from stars tidally disrupted by 
central black holes (BHs) leading to expected time-dependent apparent variabilities. And based on long-term 
variability properties, and there are more than 80 TDEs detected and reported (see detailed information in 
\url{https://tde.space/}), strongly supporting TDEs as better indicators to massive BHs and corresponding 
BH accreting systems. More recent reviews on TDEs can be found in \citet{st18}. Based on well improved 
theoretical TDEs in \citet{gr13, gm14}, accreting fallback debris in TDEs have apparent time-dependent 
structure evolutions, also leading to structure evolutions of broad emission line regions (BLRs) built 
from accreting TDE debris. 

    Besides direct results on both observational and theoretical TDEs, there is one special kind of broad 
line AGN (Active Galactic Nuclei), AGN with double-peaked broad low-ionization emission lines (double-peaked 
AGN) \citep{chf89, el95, st03, fe08, zh17}, with accreting debris in TDEs as one probable origination of 
emission materials of double-peaked broad emission lines (see discussions in \citet{el95}, also see simulated 
results on BLRs from TDE debris in \citet{gm14}). Therefore, if observed broad emission lines were related 
to TDEs, broad emission lines could be explained by reasonable disk-like BLRs related to TDEs, which is the 
main objective of the Letter.  

    More recently, among the reported TDEs candidates, \citet{lz17} have reported a relativistic elliptical 
accretion model well applied to describe three-month variabilities of the observed double-peaked broad H$\alpha$ in 
PTF09djl. \citet{ht19} have shown a non-axisymmetric accretion disk model applied to describe two-month 
variabilities of the observed double-peaked broad H$\alpha$ in PS18kh. \citet{sn20} have shown that the 
seven-month variabilities of broad Balmer lines have an asymmetric double-peaked phase which is consistent 
with accretion disc emissions in AT 2018hyz. \citet{hf20} have also shown that the distinct double-peaked 
broad H$\alpha$ can be well described by a low eccentricity accretion disk emissions in the 
AT 2018hyz. Meanwhile, \citet{ht16} have shown four-month variabilities of asymmetric broad H$\alpha$ 
(apparently asymmetric features shown in Fig.~8\ in \citet{ht16}) in ASASSN-14li. Therefore, accreting debris 
in TDEs can lead to double-peaked broad emission features related to disk-like emission regions. And apparently, 
the double-peaked features reported above are transient features that survive for less than 1 year, basically 
consistent with durations of TDEs determined by central BH masses around $10^6{\rm M_\odot}$ and parameters 
of disrupted solar-like main-sequence stars. Different parameters in TDEs, such as different ratios of the 
tidal radius to the pericenter distance, different central BH mass and stellar properties, etc., could lead 
to expected double-peaked features persisted for longer durations, such as the case in the TDE candidate SDSS 
J0159 with ten-year variability of asymmetric broad H$\alpha$.

\begin{figure*}
\centering\includegraphics[width = 18cm,height=6cm]{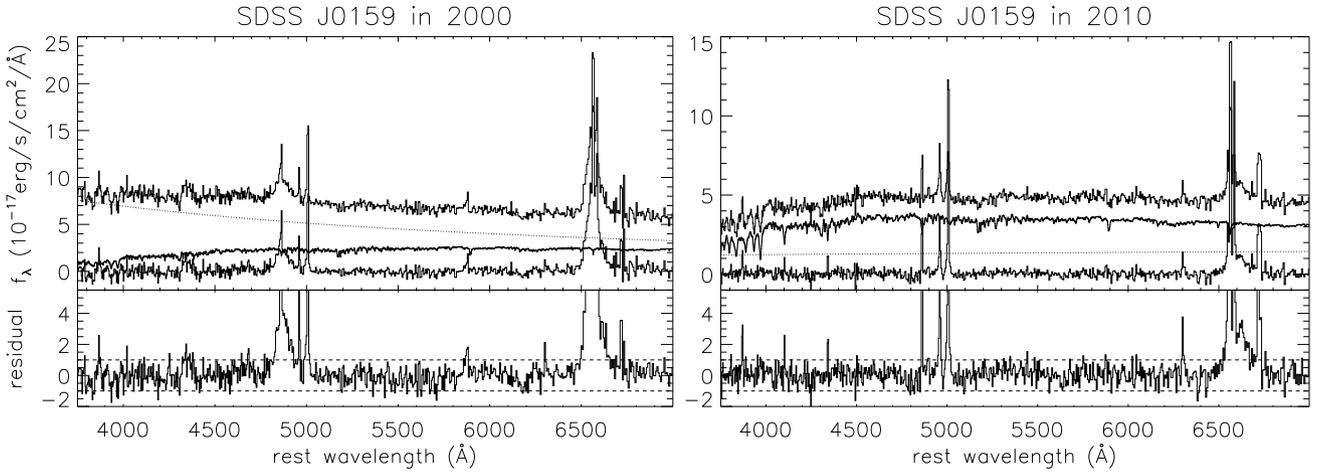}
\caption{On the determined stellar lights by the SSP method (top panels) and the corresponding residuals 
(bottom panels) in the SDSS spectrum in 2000 (left panels) and in 2010 (right panels). In each top panel, 
from top to bottom, thin solid line, thick dotted line, thick solid line and thin solid line represent 
the observed spectrum, the determined power law AGN continuum emissions, the determined stellar lights 
and the pure line spectrum, respectively. In each bottom panel, thick horizontal dashed lines show 
$residual~=~\pm1$.}
\label{spec}
\end{figure*}

   \citet{md15} have shown the observed broad H$\alpha$ in SDSS J0159 (=J015957.64+003310.5) in 2000 and 
2010, and reported interesting variabilities in ten years: broad H$\alpha$ in 2010 with FWHM (Full Width 
at Half Maximum) about $6167{\rm km/s}$ wider than that in 2000 with FWHM about $3408{\rm km/s}$ and 
central wavelength in 2000 with $\lambda_0$ about 6560\AA\ bluer than that in 2010 with $\lambda_0$ about 
6582\AA. The special variability properties of broad H$\alpha$ in SDSS J0159, especially the different 
central wavelengths, can not be clearly explained by common variabilities of broad emission lines coming 
from normal BLRs in normal AGN, however, accretion disk origination of broad H$\alpha$ related to TDEs 
could provide reasonable explanations to the observed broad H$\alpha$ variabilities in SDSS J0159. Meanwhile, 
we have recently reported properties of BH mass in SDSS J0159 in \citet{zh19} that there are different 
virial BH mass under the virialization assumption \citep{pf04} from the BH mass estimated through the 
\msig relation \citep{kh13}. The different BH masses through different methods strongly indicate that 
there should be strong contributions to broad line emission materials from central TDE debris. Therefore, 
the known changing-look AGN SDSS J0159 is a better target to discuss probable connections between its 
ten-year long variabilities of asymmetric broad H$\alpha$ and a central TDE. Section 2 presents our main 
results on variabilities of broad H$\alpha$ of SDSS J0159, and necessary discussions. Section 3 gives our 
final conclusions.

\begin{figure*}
\centering\includegraphics[width = 18cm,height=6cm]{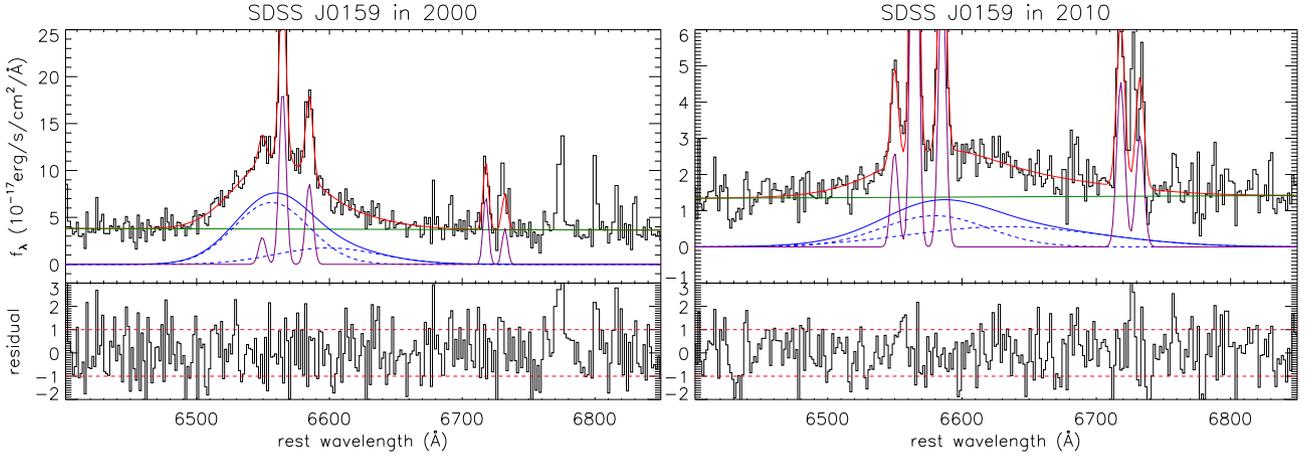}
\caption{The best fitted results (top panels) and the corresponding residuals (bottom panels) to the 
emission lines around H$\alpha$ in 2000 (left panels) and in 2010 (right panels) by multiple Gaussian functions, 
after subtractions of the stellar lights. In each top panel, black solid line and red solid line show the 
line spectrum and the best fitted results, dark green solid line shows the determined power law AGN continuum 
emissions, blue solid line shows the determined broad H$\alpha$, purple solid lines show the determined narrow 
H$\alpha$, [N~{\sc ii}] and [S~{\sc ii}] doublets, and blue dashed lines show the two Broad Gaussian components 
included in the broad H$\alpha$. In each bottom panel, red dashed lines show $residual~=~\pm1$.}
\label{line}
\end{figure*}

\section{Broad H$\alpha$ Variability in SDSS J0159} 

    Spectra of SDSS J0159 in 2000 and 2010 have been collected by PLATE-MJD-FIBERID of 0403-51871-0549 
and 3609-55201-0524 from SDSS. Then, the following steps have been applied to subtract contributions of 
both stellar lights and narrow emission lines in the SDSS spectra, in order to obtain the pure broad H$\alpha$. 
Here, the widely accepted SSP method (Simple Stellar Population) \citep{bc03, ka03} has been applied to 
determine contributions of stellar lights in the observed spectrum with emission lines being masked out, 
considering broadened stellar templates plus one power-law component for the AGN continuum emissions. 
Meanwhile, we are very familiar with the SSP method, such as what we have recently done in \citet{zh19}, 
and do not show further discussions on the SSP method any more. Fig.~\ref{spec} shows the SSP method 
determined stellar lights through the Levenberg-Marquardt least-squares minimization technique, and the 
corresponding residuals of $\frac{y-y_{\rm m}}{y_{\rm err}}$ where $y$, $y_{\rm m}$ and $y_{\rm err}$ 
represent the observed spectrum after emission lines being masked out, the determined best fitted results 
and the corresponding errors of $y$, respectively. Similar results on subtractions of the stellar lights 
can be found for SDSS J0159 in \citet{ld15}.

   After subtractions of the stellar lights, multi-Gaussian components have been applied to describe the 
emission lines around H$\alpha$ within rest wavelength from 6400\AA\ to 6850\AA. There are two broad Gaussian 
components applied to describe the broad H$\alpha$, and five another narrow Gaussian components applied to 
describe the other narrow emission lines of narrow H$\alpha$, [N~{\sc ii}] and [S~{\sc ii}] doublets. Similar 
emission line fitting procedure has been applied in our more recent paper \citet{zh17b}. Fig.~\ref{line} 
shows the best fitted results and corresponding residuals to the emission lines around H$\alpha$ through 
the Levenberg-Marquardt least-squares minimization technique. The determined $\chi^2=SSR/Dof$ values 
($SSR$ and $Dof$ as summed squared residuals and degree of freedom) are about $\sim1.31$ and $\sim1.08$ 
for the best fitted results to the emission lines in 2000 and in 2010, respectively. 
Table~1 lists the determined line parameters of broad and narrow emission lines. Here, $\lambda_0$ (first 
moment) and $\sigma$ (second moment) of the broad H$\alpha$ are determined by 
\begin{equation}
\lambda_0 = \frac{\int\lambda\times f_\lambda d\lambda}
	{\int f_\lambda d\lambda}\ \ \ \ \ \ \ \ \ \ \ \ \ \ \ \ \ \  
\sigma^2 =\frac{\int\lambda^2\times f_\lambda d\lambda}
        {\int f_\lambda d\lambda} - \lambda_0^2
\end{equation},
where $f_\lambda$ represents line profile of the broad H$\alpha$. The uncertainties of $\lambda_0$ and 
$\sigma$ and line flux are determined by the bootstrap method.

\begin{figure*}
\centering\includegraphics[width = 18cm,height=6cm]{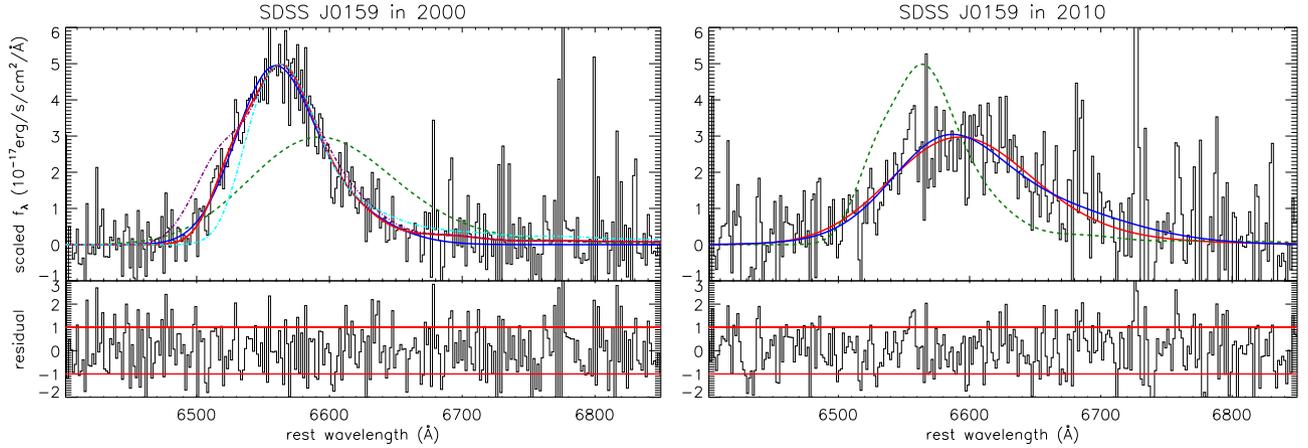}
\caption{The best fitted results (top panels) and the corresponding residuals (bottom panels) to the broad 
H$\alpha$ in 2000 (left panels) and in 2010 (right panels) by the relativistic elliptical accretion disk 
model, after subtractions of both the narrow emission lines and the power law AGN continuum emissions 
from the line spectra shown in Fig.~\ref{line}. In each top panel, black solid line and red solid line 
represent the pure broad H$\alpha$ and the best fitted results by the accretion disk model, dark green dashed 
line shows the determined best fitted results to the broad H$\alpha$ observed in the other epoch by the 
accretion disk model, blue solid line shows the scaled Gaussian fitted results to the broad H$\alpha$ shown 
in Fig.~\ref{line}. In the left panel, dot-dashed line in purple (in cyan) shows the model created line 
profile with $\sin(i)=0.196+3\times0.011$ ($e_1=0.93-3\times0.03$), but with the other parameters set to 
the model determined values. In each bottom panel, horizontal red solid lines show $residual~=~\pm1$.}
\label{model}
\end{figure*}

    The redder and broader broad H$\alpha$ in 2010 can be re-confirmed, as reported in \citet{md15}. 
Meanwhile, Table~1 also lists the information of the two broad Gaussian components shown as blue 
dashed lines in Fig.~\ref{line}. There are one blue component plus one red component in 2000, but two 
red components with much different line widths in 2010, strongly indicating that it is unreasonable 
to consider rotating independent double emission regions to explain the two Gaussian components. 
Therefore, the main objective of two Gaussian components to the broad H$\alpha$ is only to determine 
the best fitted results to the emission lines around H$\alpha$ (especially to the narrow emission lines), 
and the asymmetric line profiles of 
broad H$\alpha$ shown in Fig.~\ref{model} can be well determined after subtractions of both the narrow 
emission lines and the power-law continuum emissions from the line spectra shown in Fig.~\ref{line}. 
Then, the following accretion disk model is mainly considered.

\begin{table}
\caption{Line parameters of emission lines}
\begin{tabular}{lccc}
\hline\hline
\multicolumn{4}{c}{Emission lines in 2000} \\
\hline
Line    &   $\lambda_0$  &   $\sigma$  & flux  \\
	&    \AA   & \AA  & ${\rm 10^{-17}erg/s/cm^2}$\\
\hline
Broad H$\alpha$ & {\bf 6568.91$\pm$6.45} & {\bf 38.12$\pm$3.26} & 654$\pm$64 \\
	1st Broad  &  6556.41$\pm$10.01 & 28.21$\pm$8.19 & 460$^{*}$ \\
	2nd Broad  &  6598.64$\pm$10.63 & 39.24$\pm$11.31 & 170$^{*}$ \\
Narrow H$\alpha$ & 6564.74$\pm$0.12 & 2.57$\pm$0.13 & 120$\pm$6 \\
\nii & 6584.85$\pm$0.24 & 2.71$\pm$0.28 & 58$\pm$7 \\
\si & 6717.82$\pm$0.22 & 2.01$\pm$0.22 & 37$\pm$4 \\
\sii & 6732.21$\pm$0.23 & 2.01$\pm$0.22 & 19$\pm$5 \\
\hline
\multicolumn{4}{c}{Emission lines in 2010} \\
\hline
Broad H$\alpha$ & {\bf 6595.88$\pm$3.52} & {\bf 56.29$\pm$4.52} & 168$\pm$15 \\
	1st Broad  &  6579.73$\pm$8.63 & 39.54$\pm$14.19 &  85$^{*}$ \\
	2nd Broad  &  6632.25$\pm$8.74 & 71.04$\pm$25.73 &  100$^{*}$ \\
Narrow H$\alpha$ & 6564.71$\pm$0.06 & 2.53$\pm$0.05 & 117$\pm$3 \\
\nii &  6585.23$\pm$0.12 & 2.79$\pm$0.12 & 56$\pm$3 \\
\si &  6718.17$\pm$0.17 & 3.10$\pm$0.16 & 36$\pm$2 \\
\sii &  6732.55$\pm$0.17 & 3.11$\pm$0.16 & 25$\pm$3 \\
\hline
\end{tabular}\\
Notice: The second, third and fourth column show the central wavelength in unit of \AA\ in rest frame, 
the line width (second moment) in unit of \AA\ and the line flux in unit of $10^{-17}{\rm erg/s/cm^2}$. 
"1st Broad" and "2nd Broad" shows the parameters of the two broad Gaussian components included in 
the broad H$\alpha$. The symbol $^{*}$ means the determined uncertainties larger than the measured 
parameter by the least-squares minimization technique, due to lower spectral quality.
\end{table}

     Different kinds of relativistic accretion disk models can be found in the literature. 
Circular accretion disk model has been firstly proposed in \citet{chf89}, and then the improved elliptical
accretion disk model can be found in \citet{el95}. Besides the circular and elliptical accretion disk models, 
a model of circular disk with spiral arms can be found in \citet{sn03}, a warped accretion disk model can be found in 
\citet{hb00}, and a stochastically perturbed accretion disk model can be found in \citet{fe08}. Here, the 
elliptical accretion disk model is preferred, because the model with less number of necessary model parameters 
can explain almost all observational double-peaked features. Moreover, as the shown smooth profiles of asymmetric 
broad H$\alpha$ in SDSS J0159, it is not necessary to consider existences of arms and/or warped structures 
and/or bright spots which are mainly applied to subtle structures of broad line profiles (such as some cusps 
around peaks/shoulders). Detailed descriptions of the relativistic elliptical accretion disk model with seven 
model parameters can be found in \citet{el95}. Meanwhile, we have also applied the very familiar elliptical 
accretion disk model, see our studies on double-peaked lines in \citet{zh05, zh13, zh15}.

    There are seven necessary model parameters ($par$): inner boundary $r_0$, out boundary $r_1$, inclination 
angle $i$ of disk-like BLRs, eccentricity $e$ and orientation angle $\phi_0$ of elliptical rings, local broadening 
velocity $\sigma_L$, line emissivity slope $q$ ($f_r~\propto~r^{-q}$). Then, the relativistic elliptical accretion 
disk model with $r_0$ well extended down to $\sim6{\rm R_G}$ can be applied to {\bf simultaneously} describe the 
asymmetric broad H$\alpha$ in 2000 and in 2010 as follows. The line profiles of broad H$\alpha$ in 2000 and in 
2010 are [$\lambda_1$, $f_{\lambda,~1}$], [$\lambda_{2}$, $f_{\lambda,~2}$], 
respectively, leading to a combined line profile as [$\lambda$,~$f_\lambda$] with $\lambda=[\lambda_1,~\lambda_2]$ 
and $f_\lambda=[f_{\lambda,~1},~f_{\lambda,~2}]$. Then, the model expected line profile can be written as 
[$\lambda_M$,~$f_{\lambda,~M}$] with $\lambda_M=[\lambda_1,~\lambda_2]$ and $f_{\lambda,~M}=[M_{par1},~M_{par2}]$. 
Each model determined line profile of $M_{par1}$ and $M_{par2}$ has seven model parameters, however, due to the 
accepted almost same regions and same inclination angle of the disk-like BLRs for the broad H$\alpha$ in 2000 and 
in 2010, there are finally 11 model parameters as follows: $r_0$, $r_1$ and $i$ for the inner and outer boundaries 
and the inclination angle of the disk-like BLRs, $q_1$ and $q_2$ as the line emissivity slopes for the broad 
H$\alpha$ in 2000 and in 2010, $\sigma_{L,~1}$ and $\sigma_{L,~2}$ as the local broadening velocities for the broad 
H$\alpha$ in 2000 and in 2010, $e_1$, $e_2$ and $\phi_{0,~1}$ and $\phi_{0,~2}$ as the eccentricities and orientation 
angles for the disk-like BLRs of the broad H$\alpha$ in 2000 and in 2010. Meanwhile, due to weaker broad H$\alpha$ 
in 2010 as shown in Fig.~\ref{line}, mean line intensities of pure broad H$\alpha$ in 2000 and  2010 are applied 
to scale the broad H$\alpha$ in 2000 and 2010, leading to the scaled broad H$\alpha$ in 2000 and 2010 having similar 
line intensities. Then, through the Levenberg-Marquardt least-squares minimization technique, the best fitted 
results with $\chi^2\sim1.21$ and the corresponding residuals to the broad H$\alpha$ can be well 
determined and shown in Fig.~\ref{model}, strongly indicating the relativistic elliptical accretion disk model 
is an appropriate model.

   The model parameters are determined as $r_{0}\sim10.5\pm1.1{\rm R_G}$, $r_{1}\sim129.6\pm31.7{\rm R_G}$\footnote{Even 
different $r_1$ in model functions applied to emission regions of broad H$\alpha$ in 2000 and in 2010, similar $r_1$ 
can be found, but with larger uncertainties.}, $\sin(i)\sim0.196\pm0.011$ (about 11.3\degr), $q_{1}\sim1.82\pm0.19$, 
$q_2\sim0.72\pm2.13$\footnote{Probably due to lower spectral quality of broad H$\alpha$ in 2010, the determined model 
parameters of $q_2$ and $\phi_{0,~2}$ have their model determined values smaller than their corresponding uncertainties.}, 
$\sigma_{L,~1}\sim730\pm96{\rm km/s}$, $\sigma_{L,~2}\sim2230\pm762{\rm km/s}$, $e_1\sim0.93\pm0.03$, $e_2\sim0.88\pm0.46$, 
$\phi_{0,~1}\sim-31\pm2\degr$ and $\phi_{0,~2}\sim44\pm60\degr$. In addition, the model determined disk-like BLRs with 
a small inclination angle (nearly face on) reasonably lead to the observed single-peaked asymmetric broad H$\alpha$ in 
SDSS J0159. Moreover, among the determined model parameters, $\sin(i)$ and $e_1$ have quite smaller uncertainties, 
mainly due to more sensitively dependence on $\sin(i)$ and $e_1$, In order to show further information on the tightly 
constrained parameters of $\sin(i)$ and $e_1$, two additional model created profiles have been shown in the left panel 
of Fig.~\ref{model}: one is created through the elliptical accretion disk model applied with the value of $\sin(i)$ to 
be $\sin(i)=0.196+3\times0.011$ (where 0.011 is the determined uncertainty of $\sin(i)$) and with the other parameters 
set to the model determined values, the other one is created with the value of $e_1$ to be $e_1=0.93-3\times0.03$ (where 
0.03 is the determined uncertainty of $e_1$) and with the other parameters to the model determined values. It is clear 
that smaller deviations from the model determined values of $\sin(i)$ and $e_1$ could lead to the worse model determined 
profiles to describe the asymmetric broad H$\alpha$. Therefore, the smaller uncertainties of $\sin(i)$ and $e_1$ can 
be well accepted. Meanwhile, we can find that the model determined parameters are different from the statistical results 
on disk-like BLRs in a large sample of double-peaked SDSS AGN in \citet{st03}, especially the smaller inner and outer 
boundaries and larger eccentricity which will instead provide further clues on special properties of the disk-like 
BLRs in SDSS J0159. 

      Before proceeding further, there is one interesting point we should note. In the Letter, the applied 
relativistic elliptical accretion disk model is the one well discussed in \citet{el95}, a bit different from 
the model discussed in \citet{lz17}, such as the parameter of line emissivity slope as a free model parameter 
in our applied model but fixed in the model discussed in \citet{lz17}.Actually, if the parameter of $q$ 
is fixed to 3\ in the applied accretion disk model, the re-determined best fitted results could lead to 
$\chi^2\sim1.58$ larger than $\chi^2\sim1.21$ for the model with $q$ as a free model parameter, indicating 
dependence of the results on $q$. Therefore, $q$ as a free model parameter can be preferred and lead to 
more appropriate final best fitted results. And, it is very interesting to find 
that the model determined eccentricity is large and inner boundary is small for the disk-like BLRs in SDSS 
J0159, but similar as the reported results in \citet{lz17} with $e\sim0.96$ and $r_0\sim11{\rm R_G}$ for 
the disk-like broad line emission regions in PTF09djl. However, there are larger inner boundaries and 
smaller eccentricities of the determined disk-like regions with $r_0\sim1000{\rm R_G}$ and $e\sim0.1$ in 
AT 2018hyz and with $r_0~\sim~60{\rm R_G}$ and $e\sim0.25$ in PS18kh. Comparing the disk parameters in the TDEs 
candidates indicate the preferred elliptical accretion disk-like BLRs related to a central TDE in SDSS J0159 
similar as the case in PTF09djl, but quite different from the cases in AT 2018hyz and PS18kh, mainly due to 
different geometric structures from TDEs with different physical parameters.

     Moreover, before giving the necessary discussions, we give some detailed arguments on the important 
parameter of BH mass in SDSS J0159. As discussed in \citet{md15, ld15}, the virial BH mass about 
$10^8{\rm M_\odot}$ has been reported. However, \citet{zh19} and \citet{we19} have shown that the 
stellar velocity dispersion in SDSS J0159 is smaller than the expected value from the virial BH mass 
through the \msig relation. If the smaller stellar velocity dispersions were accepted that 
$\sigma\sim80{\rm km/s}$ in \citet{zh19} and $\sigma\sim120{\rm km/s}$ in \citet{we19}, the BH mass 
from the \msig relation discussed in \citet{zh19} could be around $2.1\times10^6{\rm M_\odot}$ and 
$1.4\times10^7{\rm M_\odot}$, respectively. After considering the strong contributions of TDE debris to 
broad lines, rather than the virial BH mass, the mean BH mass about $8\times10^6{\rm M_\odot}$ (smaller 
than the virial BH mass) has been adopted through the \msig relation. Then, based on the determined 
model parameters in the relativistic elliptical accretion disk model, there are at least two major 
points we can confirm. 

    First, if the disk-like BLRs in SDSS J0159 was based on accreting materials from a central TDE, 
we would expect that the inner boundary of the disk-like BLRs could be near to (at least not very different 
from) the corresponding tidal disruption radius $R_{\rm TDE}$. For TDEs around Schwarzchild BHs, tidal disruption 
radius $R_{\rm TDE}$ is about $R_{\rm TDE}\sim10R_{G}\times(\frac{M_\star}{\rm M_\odot})^{-1/3}(\frac{M_{\rm BH}}
{\rm 10^7 M_\odot})^{-2/3}\frac{R_\star}{\rm R_\odot}$. For SDSS J0159, \citet{md15} have shown that the 
central tidally disrupted star was a main-sequence star with stellar mass about (or larger) 1.2${\rm M_\odot}$ 
with the virial BH mass $10^8{\rm M_\odot}$ accepted. However, through the \msig relation, the expected BH 
mass is smaller than the virial BH mass, therefore, the determined stellar mass should be smaller than than 
the value of 1.2${\rm M_\odot}$. Then, with $M_{\rm BH}\sim8\times10^6{\rm M_\odot}$ and $\frac{M_\star}{\rm 
M_\odot}<1.2$, we would have $R_{\rm TDE}<13{\rm R_G}$ similar as the model determined $r_0$, strongly 
indicating that the model determined small inner boundary $r_0$ is reasonable enough, and also supporting that 
materials in the disk-like BLRs could be tightly related to accreting debris in a central TDE. Moreover, 
through the criterion that $R_{\rm TDE}>2{\rm R_G}$, we also have the lower limit for the stellar mass is 
about $0.05{\rm M_\odot}$. The following detailed descriptions to the long-term light curve by TDE models 
could be applied to check whether our expected stellar mass is valid.

     Second, we can confirm the determined disk-like BLRs in SDSS J0159 are not similar as common BLRs 
in normal broad line AGN. Based on reported BLRs sizes of the reverberation mapped AGN, BLRs sizes of 
normal broad line AGN can be well estimated through the well-known R-L empirical relation \citep{bd13} by 
continuum and/or broad line luminosity. If the R-L relation was applied, the BLRs size was about 
30~light-days in SDSS J0159 reported in \citet{ld15}. However, based on the model determined boundaries 
of disk-like BLRs in SDSS J0159, the BLRs size of SDSS J0159 was about $75{\rm R_G}$ (the flux-weighted radius) 
leading to 0.04~light-days with the BH mass about $8\times10^6{\rm M_\odot}$ accepted. Therefore, the BLRs 
for the observed broad H$\alpha$ of SDSS J0159 through the relativistic elliptical accretion disk model 
are different from the commonly expected BLRs in normal broad line AGN. The results strongly support 
the assumption that the broad H$\alpha$ emission materials in SDSS J0159 are from TDE debris nearer 
to central BH.

\section{Conclusions}

   Finally, we give our main conclusions as follows. We can find that the relativistic elliptical disk 
model determined disk-like BLRs not only can be applied to well explain the asymmetric broad H$\alpha$ 
variabilities in SDSS J0159 indicating the disk-like BLRs are preferred, but also the determined disk-like 
BLRs have locations similar to the regions of accreting debris in a TDE indicating tight connections 
between the disk-like BLRs and TDE debris. Therefore, the well explained broad H$\alpha$ variability 
properties through the expected accretion disk origination actually provide further evidence to support 
a central TDE in the known changing-look AGN SDSS J0159. 

\section*{Acknowledgements}
Zhang gratefully acknowledge the anonymous referee for giving us constructive comments and suggestions 
to greatly improve our paper. Zhang gratefully acknowledges the kind support of Starting Research Fund 
of Nanjing Normal University and from the financial support of NSFC-11973029. This Letter has made 
use of the data from the SDSS projects. The SDSS-III web site is http://www.sdss3.org/. SDSS-III is 
managed by the Astrophysical Research Consortium for the Participating Institutions of the SDSS-III Collaboration.

\section*{Data Availability}
The data underlying this article will be shared on reasonable request to the corresponding 
author (\href{mailto:xgzhang@njnu.edu.cn}{xgzhang@njnu.edu.cn}).

\label{lastpage}
\end{document}